# AUTHORISED TRANSLATIONS OF ELECTRONIC DOCUMENTS

**Jan Piechalski and Andreas U. Schmidt**

Fraunhofer-Insitute for Secure Information Technology SIT

Rheinstrasse 75

64295 Darmstadt, Germany

Phone: +49 6151 869 60227

Fax: +49 6151 869 224

{andreas.u.schmidt,jan.piechalski}@sit.fraunhofer.de

ABSTRACT

A concept is proposed to extend authorised translations of documents to electronically signed, digital documents. Central element of the solution is an electronic seal, embodied as an XML data structure, which attests to the correctness of the translation and the authorisation of the translator. The seal contains a digital signature binding together original and translated document, thus enabling forensic inspection and therefore legal security in the appropriation of the translation. Organisational aspects of possible implementation variants of electronic authorised translations are discussed and a realisation as a stand-alone web-service is presented.

ACM CLASSIFICATION

H.4.1 [Information Systems Applications]: Office Automation – Workflow Management;

K.6.5 [Management of Computing and Information Systems]: Security and Protection – Authentication;

# AUTHORISED TRANSLATIONS OF ELECTRONIC DOCUMENTS

## 1 INTRODUCTION

Authorised translations are, today, a business entirely restricted to paper documents. Though translators use IT-tools to generate translations, the end-product is, in many jurisdictions, still a paper print-out, undersigned by the translator who places a stamp on it to attest the accuracy of the translation and to express his authorisation. Nevertheless, within the EU, the necessary legal background to model this proceeding completely using digital documents and electronic signatures is already in place for some time.

The present paper is a proposal to carry out authorised translations in an entirely electronic environment, and in a way which achieves the ultimate goal of preserving the probative force of the translated document. Starting from an analysis of requirements for authorised translations, we view and analyse authorised translations in the larger framework of general transformations, such as changes of data format of signed digital documents. Research on this subject was carried out in the project TransiDoc funded by the German ministry of Labour and Economics [1, 2]. It resulted in the central concept of a *transformation seal* which can be used as a prototype for the application-specific profile of a *translation seal.* This base concept for the realisation of electronic authorised translations incorporates the signature of the translator and relevant meta-data, and binds original and translation inseparably together. Added value can be generated for electronically authorised translations, if the original itself carries electronic signatures. While paper translations mostly carry a note 'illegible signature' in the approximate place of the original ones, here the translator can verify the digital signatures in the source and carry that verification data into the target in some form.

The paper is organised as follows. In section 2 we present a general, abstract analysis of legally secure document transformations. Section 3 describes requirements for the special case of authorised translations in the German legal domain, which turn out to be scarce and diversified. To make a comparison to a better regulated area, we take a look at certifications by (staff of) public authorities and notaries public, for which rather clear regulations exist for digital realisations using electronic signatures. This will serve us as a blueprint for the data structures for authorised translations. We spell out the (small) set of proper requirements for electronic authorised translations ensuing form the previous consideration in Section 4 and relate them to quality assurance standards for translations [3], and applicable standards to ensure interoperability of language-dependent digital objects, i.e, *internationalisation*. Section 5 describes the data structures embodying the translation seal, which are of general value beyond a particular implementation. Section 6 introduces two deployment variants, an 'ideal' one which makes use of widespread PKI and signature technology which exhibits organisational problems of general interest, and a 'down to earth' approach demonstrating authorised electronic translations in a web service, posing minimal requirements on the part of the translator using it. Section 7 contains conclusions and outlook to further work. Section 8 is an Appendix containing some technical material.

## 2 LEGALLY SECURE DOCUMENT TRANSFORMATIONS

Legal security is the prime objective in the handling of electronically signed digital documents. This regards all aspects of the life-cycle of a document, in particular its secure long-term archiving and transformations, of any kind, of it. The latter point is the focus of the research project TransiDoc, aiming at a framework of data structures, organisational guidelines and best practices for document transformations. The fundamental theoretical concepts of the TransiDoc approach have been expounded in [1]. Authorised translations are in fact a special instantiation of this general conceptual framework and thus we briefly present it here.

Essential for legal security is that the result of a transformation – the *target document* – is usable in the desired application context, i.e., unfolds the necessary probative force. To achieve this certain, application-dependent properties of the *source documen*t must preserved in the transformation and carried over to the target. To ensure probative force, the transformation process should provide ample possibilities for forensic inspection of the transformation process and its result. Viewed as a process this leads to the structuring of a general transformation (including such between electronic and paper documents) into phases as shown in Figure 1.

In an initial step called *classification* the source document is inspected to determine the purpose of the transformation. Apart from ascertaining source formats this is essential for the whole following process, since it not only determine the relevant properties of the source but also those of the target and the transformation, which have to be satisfied. The classification also determines a *rule-set* an abstract term for the comprehensive set of rules governing all the following phases. Generically, the rule-set consists of a combination of machine-processible instructions, with normative prescriptions understandable only by humans. In some cases, like notarisations, the latter can already be implicated by existing legal regulations. Application specific rule-sets are envisaged to be generated from generic ones by profiling processes. In fact the rules and data structures for translations represent such a profile.

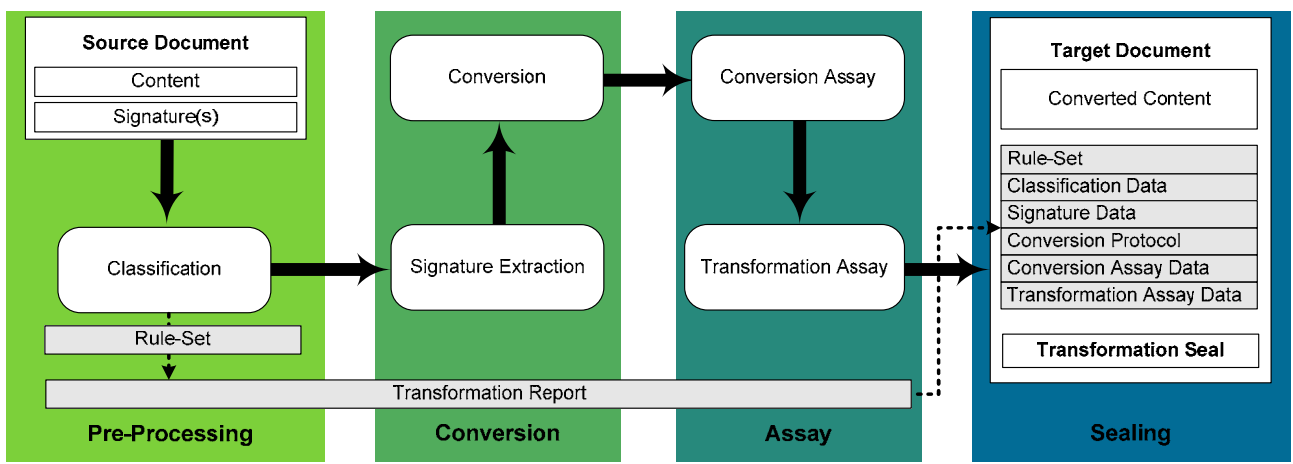

*Figure 1: Phases of a legally secure transformation process*

As a data container which carries the information compiled during the transformation, the *transformation report* is useful to establish security by conserving relevant meta data and, e.g., protocols of the conversions and inspections carried out in later phases. The first item in the transformation report is the rule-set. The transformation report serves also to ensure the proper binding of relevant data with each other, namely 1. The source contents must be uniquely identified throughout the whole process, for which the record carries an identifier. 2. Likewise, the rule-set must be unique during the process. 3. The integrity of the target's contents and their association with the source's must be ensured. 4. Protocols and meta data generated must be kept unique for the process and unadulterated.

During *signature extraction*, the signatures of the source are gathered from it and added to the transformation record as source signatures. The rule-set determines whether digital signatures must be verified and prescribes validation policies and names the signature data to be carried to the record (e.g., time stamps, attributes, etc.). In the *conversion* phase the proper conversion of source to target contents takes place according to the rules of the rule-set. Apart from the target contents, a conversion protocol and error log is filed in the report.

In many cases it is possible to include two steps of ex post inspection into the transformation process to raise the level of trustworthiness, for which we use the term *assay*. The first step assays the results of the conversion of the contents by any means possible, and as prescribed by the rule-set. This can mean anything from a person comparing source document and converted contents – as in the case of translations – to merely checking the syntactic compliance of the converted contents with a specific data format (e.g., an XML Schema). Similar checks can take place for signature data. A final assaying step called *transformation assay* can inspect the correctness of the whole transformation process. For instance in distributed transformation systems, it can be necessary to ascertain that all necessary phases have been traversed, or to counter-check the hash values associated with certain parts of the transformation record.

After the two assaying steps a transformation seal is attached to the transformed document and signed by the transforming entity. The possibility to assess the quality of a transformation a postiori is an important building block for the probative force of the target. It is embodied in three subordinate goals, which describe the essential purpose of the transformation seal. 1. Securing the integrity of the transformed document, and other recorded data. 2. Attestation of the correctness of the transformation according to the specified rule set. 3. Attribution of the transformation to the transforming entity and non-repudiation of that fact. Technically, the transformation seal can be realised as a cryptographically secured, e.g., electronically signed, data container and selected data from the transformation record and other relevant meta data.

## 3   ATTESTATIONS AND TRANSLATIONS IN THE GERMAN LEGAL DOMAIN

### 3.1   Authorised Translations[1]

Under German law, an authorised translation is performed by a professional translator, sworn, registered with a judicial circuit corresponding to the location of his office, and equipped with a special seal for the purpose of translations. Although legal regulations concerning authorising of translators vary slightly within Germany between federal states, district courts (*Landgerichte*) are responsible for authorisations in most cases. Before being able to apply for authorisation, a translator is required to pass a state exam.

German law does not make any restrictions to the way authorised translations are created (except the content of the "certificate of accuracy" that is specified by numerous federal state laws and/or by-laws). However, there are standard procedures commonly accepted by translators. A well-known reference used by translators is the "Bulletin for creating authorised translations" issued by the Hamburg Department of the Interior [4]. The German standard DIN 2345 [3], also applied by translators, defines rules for quality assurance in translations.

Most of these guidelines can be applied both to paper-based and electronic translations. We only mention guidelines that have to be adapted to the process of creating translations of electronic documents. 1. The translator should check the source document for illegibility or other severe defects that would forbid performing a translation. Distinctive features within the source document (e.g. corrections or cancellations) should be mentioned in an *annotation*. 2. Handwritten signatures in the source document should be described in the target document, if inevitable as "illegible". 3. The translation should end with a "certificate of accuracy" which consists of a statement like "I hereby certify the completeness and accuracy of the translation", place, date, stamp of the seal and signature. It also can contain annotations made by the translator (e.g. transliterations of names or calendar transformations having been applied). 4. It is common practice to bind the source document and the target document together to make them inseparable as items of probative force by stapling the source with the target and stamping the seal over the staple.

---

[1] We use the term ``authorised translation'', as opposed to, e.g., ``certified translation'' to distinguish it clearly from certifications or notarisations that are performed by a notary public. These two kinds of certifications have a quite different meaning under German law.

## 3.2 Attestations of Electronic Documents

In contrast to the rather scarce and loose regulations for authorised translations, there are legal regulations concerning the performing of certifications, i.e. certifications of a certificate's copy, including detailed instructions about the content of a certification. German law differentiates between official certifications, done by state's authorities [5], and public certifications [6], done by notary publics. In both cases the law is not restricted to paper documents only, but it also allows certifications of electronic, digitally signed documents.

Although these legal regulations cannot be applied directly to authorised translations, there are obvious parallels between authorised translations and certifications of a document's copy (or rather transformations, since literal copies of digital documents make little sense, see [1]). Thus some standards developed for certifications of electronic documents can be used in the process of translating electronic documents.

Especially a certification's annotation according to § 33 *Verwaltungsverfahrensgesetz* [5] and §§39, 39a and 42 *Beurkundungsgesetz* [6] is similar to an annotation required for translations. In particular, it has to contain following elements:

1. Name of the source document
2. An optional indication that the usage of the certified target document is restricted to a certain authority
3. An optional description of defects found in the original document
4. Information about original signatures, including validation results
5. Attestation that the content of the source and the target are identical
6. Creation time of the certification
7. Creation place of the certification
8. A qualified signature of a person authorised to perform the certification.

We tentatively employ the requirements for electronic certifications as a sufficient set to guarantee legal security for the case of authorised translations.

## 4   REQUIREMENTS FOR ELECTRONIC TRANSLATIONS

A counterpart for authorised translations is missing for digital, electronically signed documents, although the necessary legal background to model this proceeding already exists within the EU. As we mentioned in Section 3.1, some of the standard procedures used while creating paper-based translations need to be adapted to the electronic process. Most significant difference is the treatment of signatures. While translating paper documents, signer authentication hardly ever takes place and is reduced to a simple description like "illegible signature" in most cases. In contrast to that, signature extraction for electronic translations offers much more possibilities, provided the PKIs of the source and the target country are interoperable. Firstly, the translator can extract digital signatures in the source, verify signer's identity and carry the results into the target. Secondly, the digital signature allows the translator verifying the integrity of the source, so he can be sure the content of the source has not been manipulated. The question of integrity is related to the presentation problem of electronic documents – the translator must be sure to be shown all signed contents. Thus he has to use a trusted viewer for the signed document format and he must have access to the PKI in the source country that was used to create the source's signatures.

In analogy to paper documents which are signed and sealed, two authentication characteristics will generally be required for a legally secure seal. An electronic signature identifying the transforming person (or entity, where such is admissible), and a means to authenticate his/her role as a person authorised to carry out the transformation. The use of attribute certificates in translation seals is a possible solution. Such attribute certificates can be issued by any CA, but the issue must be supervised by the authorities issuing seals for authorised translators (district court) in order to meet legal requirements. Furthermore, the district court must be able to revoke the attribute certificate, e. g. when the translator moves his office to another district.

Finally, it should be possible to bind inseparably the source document and the target document as it is often done with paper documents. This can be solved easily by inserting the entire source document into the annotation field of the target. After the target has been digitally signed, it is no longer possible to replace or change the source content.

Among elements of a certification's annotation listed in section 3.2, elements 1 seems superfluous, since documents to be translated generically carry a title as part of the content to be translated, and thus in particular the target's title is only determined in the transformation process. Element 2 is only applicable to official certifications. The remaining elements 3 – 8 are, in the following, adapted to translations and contained in the annotation.

## 4.1 Applicable internationalisation standards

Internationalisation for interoperable IT applications is a nontrivial task, which also our approach to electronic authorised translations has to face[2]. One key point is the necessary inclusion of source and target language in the annotation. Ethnologue [7] is an effort to list all spoken human languages, on which other standards rest in particular ISO 639-3. The aim of that standard is to enable the uniform identification of all known human languages in information systems. ISO 639-3 was devised to enable the uniform identification of all known languages in a wide range of applications, particularly including information systems. It provides as complete an enumeration of languages as possible, including living, extinct, ancient, and constructed languages, whether major or minor, and contains the Ethnologue list as a subset.

The ISO standards that can be applied for language specification in electronic translations are the ISO 639-1 and the more comprehensive ISO 639-2 [8] standards, which offer well-defined codes to describe languages (e.g. the code for English is 'en' in ISO 639-1 and 'eng' in ISO 639-2). The use of language codes for internet applications is specified in the IETF RFC 3066 in an extensible manner, based on ISO 639-2. The XML Schema [9] type 'xs:language' applies RFC 3066 and we use this type to specify the source and the target language of a translation in the annotation (see figure 4). It is anticipated that ISO 639-3 will be used in the future in these standards as well.

Transcription of names and titles is another difficult subject. DIN 2345 provides an overview of applicable standards for transliterations, i.e., conversion of titles, person's names, and short forms thereof. The actually used transliteration standards should be mentioned in the annotation. Table 1 in the Appendix gives an overview of transliteration standards into Latin characters. If there is no applicable transliteration standard, [3] prescribes the use of common phonetic rules for transcription of names and titles.

Date and time may also be subject to an appropriate conversion during translation. This is straightforward for paper based documents, where for instance conversions from Buddhistic (as used officially, e.g., in Thailand) to Gregorianic calendar are frequently done according to standard schemes. In digital documents however, it can more often be the case that date and time to the precision of seconds is of the essence, and might even be certified by electronic time stamps from legally trusted time services. Standards concerning notation of date and time are ISO 2015 and ISO 8601,[10], and as a reference base, Universal Coordinated Time (UTC) has become accepted internationally. Local times can differ from UTC by an offset of some seconds. See [11] for a detailed discussion. Whether it becomes necessary to convert times according to small shifts of the local time reference framework in a translation, may depend on the application context.

---

[2] We avoid problems of character encoding which we assume to be fixed at UTF-8 if the data format of source and target is plain text, or leave it to the specifics of the document format like PDF.

## 5 DATA STRUCTURES

This section elucidates the design of the central data structure 'translation seal'.

A transformation process in TransiDoc is specified by a *workflow definition* which can be realised as proposed in the TransiDoc project or in any other XML-based language like XPDL [11] or XACML [12]. A workflow definition consists of a sequence of *activities*. Each activity is specified by its *ActivityData* and can contain a list of rules to be executed. The performer of an activity can be a person (*operator*) or – in case of an automatic activity – a software or hardware *component*. The results of a transformation process are logged in a *workflow report*.

The workflow definition for the translation process consists of the activities classification, signature extraction, conversion, conversion assay and transformation assay, where both conversion (translation) and conversion assay are manual steps carried out by the translator. For a detailed description of the activities and corresponding rules see table 2 in the Appendix.

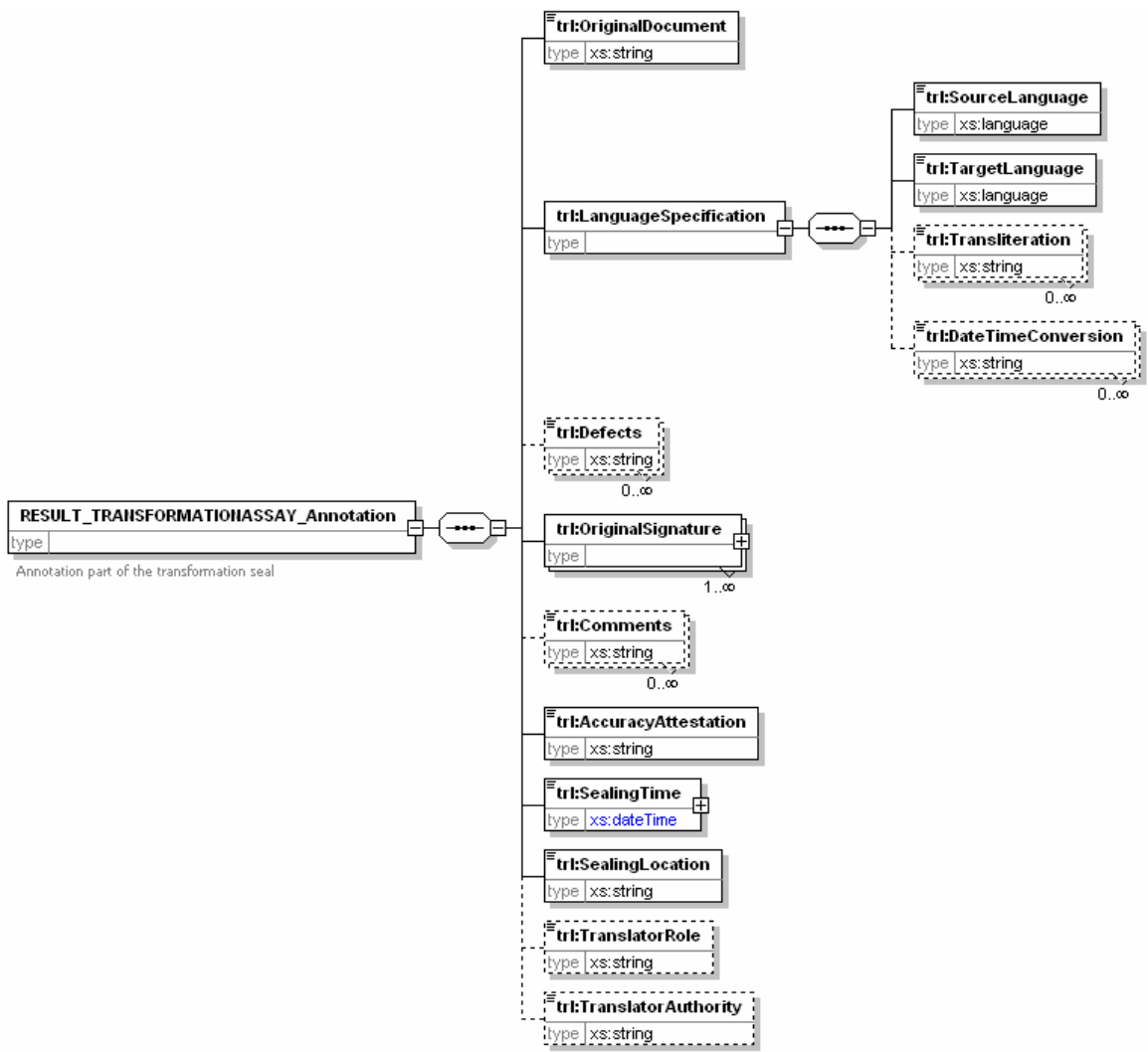

*Figure 4: Annotation of the target document*

The translation seal is a signed annotation of the translated content. An annotation that is attached to an electronic translation is similar to an annotation used for electronic certifications (see figures 4 and 5). Complete information about original signatures is stored in the workflow report during the activity "signature extraction". Since the workflow report is part of the transformation seal, this signature information is also contained there. Thus, the annotation only contains some parts of the certificate data of original signatures.

According to requirements described in Section 4, the annotation should contain data elements shown in Figure 4. Inserting the entire source document into the annotation is a method of binding source and target inseparably together to enable forensic inspection. *LanguageSpecification* contains information about languages used in the source and the target as well as methods of transliteration and calendar conversion, basing on internationalisation standards described in section 4.1. If any defects have been found in the source document during signature extraction, they should be mentioned in *Defects*. Note that the translator's name is missing in the annotation because it is part of the XML signature in the translation seal. For a detailed functional description of each element see table 3 in the Appendix.

Figure 5 shows data carried over from the original signature(s) which is part of the annotation. *OriginalSignature* contains all data necessary to report the results of signer authentication. Main elements are validation result, signer's name, signing time and information about certificates and attribute certificates. For detailed description of each element see tables 4 – 6 in the Appendix.

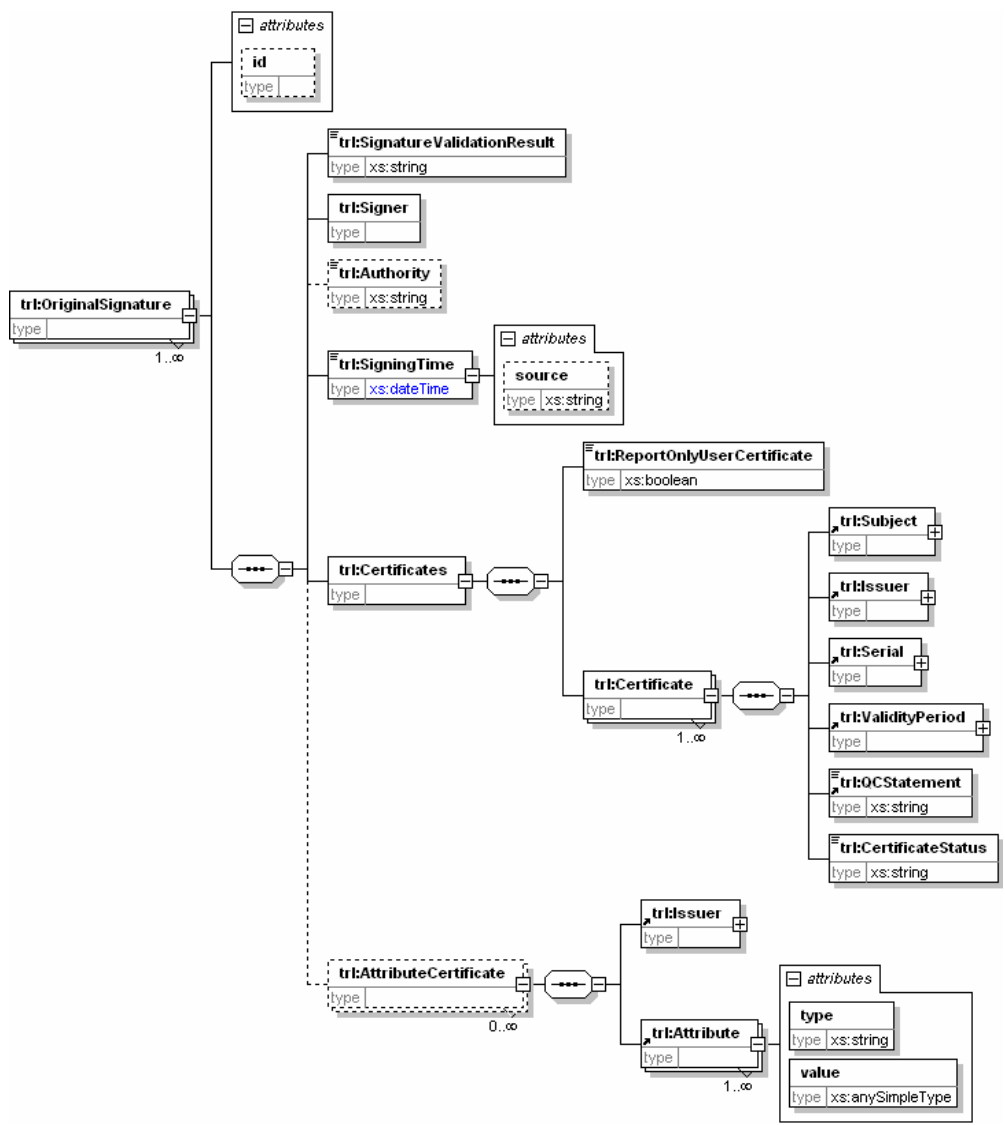

*Figure 5: Data carried over form the original signature(s)*

## 6  APPLICATION SCENARIOS AND REALISATION

We examine two different application scenarios. In the first "PKI-based" scenario the translator is required to have the ability to extract source signatures and to create his own signatures in the target document. He also must possess an attribute certificate authorising him as a translator and he has to use TransiDoc-software allowing him to create transformation seals.

In the second scenario the TransiDoc web service acts as an intermediary between the translator and the PKI. Here the translator does not have to care about extracting signatures or creating transformation seals, all this is done by the web service. The translator does not need any dedicated TransiDoc-software, or, in particular, signature creation or handling software. This is the scenario that we have actually implemented as a demonstrator.

### 6.1  PKI-based solution

Figure 2 shows the PKI-based scenario. In this figure the signature CA and the attribute certificate CA are two different entities, though it is possible that both services are provided by one CA.

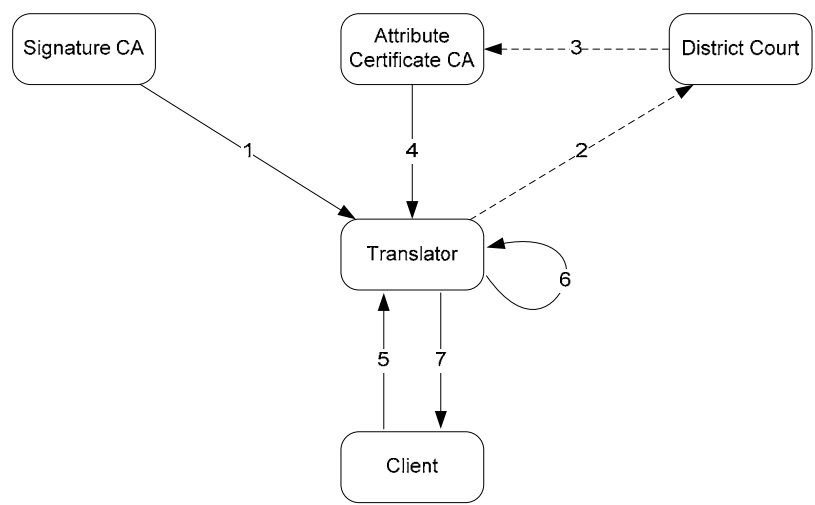

*Figure 2: PKI-world translation process*

1. The translator receives a public key, a private key and an "ordinary" certificate from the signature CA.
2. The translator applies for authorisation at the district court.
3. The district court initiates the issue of an attribute certificate by the attribute certificate CA.
4. The translator receives an attribute certificate authenticating him as an authorised translator from the attribute certificate CA.
5. The client sends a source document to the translator and orders a translation.
6. The translator performs the translation, seals the target document with a transformation seal including his attribute certificate and signs the target.
7. The translator sends the target document to the client.

This PKI-based solution is an ideal in the following sense. Deployment of technology for qualified electronic signatures is increasing in Europe and in particular in Germany – a forerunner in the creation of pertinent legal regulations, see [1]. But this holds in reality only for a limited number of professional groups. Mentionable are lawyers and notaries public, who are supported by powerful professional bodies such as the federal chamber of notaries (Bundesnotarkammer) and specialised IT support companies such as the DATEV eG. Only this professional infrastructure made it possible for policymakers to issue by-laws binding for instance notaries public to be able to exert electronic certifications using qualified electronic signatures within 2006. Another professional group which soon will be able to use electronic signatures are medical doctors which will be equipped with signature smartcards by the German public health infrastructure. All these signatures bear the benefit of implicitly identifying the signatory in his professional role.

Though hopes are that signature technology will reach a critical mass and become widespread by introduction in these groups, the infrastructural problems are still rather difficult for free professions. For this, the profession of authorised translators is a prime example. The described heterogeneity and localisation of the registration of translators impedes the solution by attribute certificates. In particular, it will be impractical for small district courts to run a dedicated CA for this sole purpose. Thus, a common PKI for the judicial sector would be a prerequisite. Furthermore, common procedures for the management and in particular the revocation, of translator's attribute certificates would be necessary, as well as a public directory (to enable clients to verify a translator's credentials).

**6.2    Stand-alone web service solution**

The problems of a 'clean' solution to the problem at hand prompt us to look for a solution which provides what is commonly called 'commercial grade security', and poses minimal requirements on the part of the translator. The use of a web service carrying most of the workload for the creation of the translation seal suggests itself. Figure 3 shows the stand-alone web service scenario.

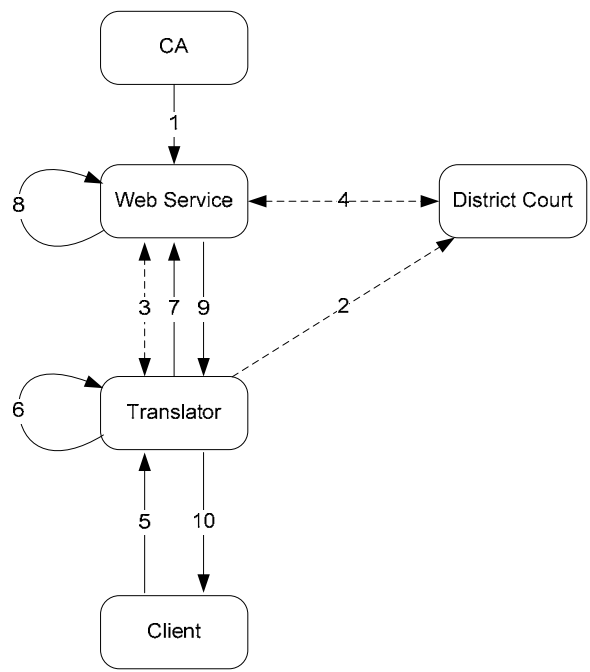

*Figure 3: Web-service solution*

1. The web service receives a public key, a private key and an "ordinary" certificate from the CA, enabling the web service to create electronic signatures.

2. The translator applies for authorisation at the district court.

3. The translator registers with the web service. In the absence of PKI and qualified signatures, this would generally be an out-of-band process involving the personal identification of the translator.

4. The web service checks the translator's authorisation at the district court and adds him to his database in case of success.

5. The client sends a source document to the translator and orders a translation.

6. The translator performs the translation.
7. The translator sends the source document and the target document to the web service.
8. The web service seals the target document with a transformation seal including his certificate and signs the target.
9. The web service sends the target document to the translator.
10. The translator sends the target document to the client. It is also possible that the client receives the target document directly from the web service.

# 7 CONCLUSIONS

We have shown that it is in principle possible to carry out purely electronic, authorised translations with a reasonable degree of legal security. This assurance is provided by binding source and target document together with other useful data using an electronic signature. The structure of the translation seal bearing these data is designed according to German law but still seems to be rather generic, so that it may be profiled to fulfil the requirements different legal domains. We have shown that it allows for structurally very different deployment variants including variants with respect to the separation of duties between stakeholders.

# 8 APPENDIX

This Appendix collects some tables with technical material.

*Table 1: Normative References for transliteration to Latin characters*

| Arabic | ISO 233 / DIN 31635 |
| Greek | ISO 843 / DIN 31634 |
| Hebrew | ISO 259 / DIN 31636 |
| Cyrillic | ISO 9 / DIN 1460 |

*Table 2: Activities and rules for the translation process*

| **Activity** | **Performer** | **Rules** |
| --- | --- | --- |
| Classification | Operator (authorised translator) + web service interface enabling the operator to enter classification data. | RULE_CLASSIFICATION_ReportOriginalDocumentClassification (The result of the classification is added to *ActivityData*) |
| | | RULE_CLASSIFICATION_CheckOriginalFormat (Only listed formats are allowed for the source document) |
| | | RULE_CLASSIFICATION_CheckTargetFormat (Only listed formats are allowed for the converted contents) |
| Signature Extraction | Web service using a component for signature verification | RULE_SIGNATUREEXTRACTION_VerifySignature (Original signatures are verified according to a specified policy) |
| | | RULE_SIGNATUREEXTRACTION_ReportSignatureData (Specifies which signature data are added to *ActivityData*) |
| Conversion | Operator | |

| Conversion Assay | Operator | |
|---|---|---|
| Transformation Assay | Operator<br>+ web service assaying the transformation automatically, showing the results, enabling the operator to enter additional information and building the annotation with data from the workflow report and with additional information entered by the operator.<br>Additional components are a component verifying signatures and a signing component | RULE_TRANSFORMATIONASSAY_CheckUsedComponents (Check if all components specified in the workflow definition have been used)<br>RULE_TRANSFORMATIONASSAY_CheckSignatureExtraction (Check if all signature data specified in the rule set are contained in the workflow report)<br>RULE_TRANSFORMATIONASSAY_CheckConsistencyOfReport (Check consistency between the workflow definition and the workflow report)<br>RULE_TRANSFORMATIONASSAY_CheckSignatures (Check signatures in the workflow report, e.g. signatures of *ActivityData*)<br>RULE_TRANSFORMATIONASSAY_CopyOriginalDocumentToAnnotation (Copy the entire original document to the annotation of the target document)<br>RULE_TRANSFORMATIONASSAY_CopyDefectsToAnnotation (Copy defects of original document to the annotation)<br>RULE_TRANSFORMATIONASSAY_CopyOrigValidationResultToAnnotation (Copy the result of signature validation to the annotation)<br>RULE_TRANSFORMATIONASSAY_CopyOrigSignatureDataToAnnotation (Specifies which signature data are copied to the annotation)<br>RULE_TRANSFORMATIONASSAY_BuildAnnotation (Specifies which additional data are added to the annotation)<br>RULE_TRANSFORMATIONASSAY_CreateSignature (Specifies the creation of the signature) |

*Table 3: Elements of the annotation*

| *OriginalDocument* | The entire original document |
|---|---|
| *LanguageSpecification* | Specification of the source language and the target language and optional specifications of the transliteration of names and the conversion method of date and time |
| *Defects* | Optional description of defects found in the original document |

| *OriginalSignature* | Information about original signatures (see figure 5 and tables 4 – 6) |
|---|---|
| *Comments* | Translator's comments |
| *AccuracyAttestation* | Attestation of accuracy and completeness of the translation |
| *SealingTime* | Creation time of the transformation seal, with an optional indication of the time source |
| *SealingLocation* | Place where the transformation seal has been created |
| *TranslatorRole* | Optional description of the role the translator is playing while performing the translation. The description of the role can be extracted from the signer's attribute certificate. |
| *TranslatorAuthority* | Optional description of the institution that authorised the translator to perform translations in the role *TranslatorRole*. The description of the institution can be extracted from the signer's attribute certificate. |

*Table 4: Data from each original signature*

| *SignatureValidationResult* | Result of the signature validation |
|---|---|
| *Signer* | Signer's name |
| *Authority* | Optional description of the authority |
| *SigningTime* | Creation time of the original signature, with an optional indication of the time source |
| *ReportOnlyUserCertificate* | Indication if the rule set requires reporting of user certificates only or all certificates of the verification path |
| *Certificates* | List of certificates (certificate data: see below) |
| *AttributeCertificates* | Optional list of attribute certificates issued for the signer's certificate (attribute certificate data: see below) |

*Table 5: Data from each original certificate*

| *Subject* | Certificate's owner with the indication of his distinguished name |
|---|---|
| *Issuer* | Certificate's issuer with the indication of his distinguished name |
| *Serial* | Certificate's serial number |
| *ValidityPeriod* | Certificate's validity period |
| *QCStatement* | Indication if the original signature is a "qualified signature" |
| *CertificateStatus* | Current status of the certificate |

*Table 6: Data from each original attribute certificate*

| *Issuer* | Attribute certificate's issuer with the indication of his distinguished name |
|---|---|
| *Attribute* | List of attributes, each defined by a type and a value |